\documentclass[journal]{IEEEtran}
\usepackage{cite,amsmath,amssymb,amsthm,amsfonts,color,bbm,graphicx,subfig}

\makeatletter
\def\endthebibliography{%
  \def\@noitemerr{\@latex@warning{Empty `thebibliography' environment}}%
  \endlist
}
\makeatother

\begin{document}
\bstctlcite{IEEE:BSTcontrol}

\title{Shannon-like Interpolation with Spectral Priors and Weighted Hilbert Spaces: Beyond the Nyquist Rate}

\author{Justin P. Haldar,~\IEEEmembership{Senior Member, ~IEEE}%
	\thanks{This work was supported in part by NIH research grants R56-EB034349 and R01-MH116173.}%
	\thanks{J. Haldar is with the Signal and Image Processing Institute, Ming Hsieh Department of Electrical and Computer Engineering, University of Southern California, Los Angeles, CA, 90089 USA (e-mail: jhaldar@usc.edu).}}

\maketitle

\begin{abstract}
In this work, we draw connections between the classical Shannon interpolation of bandlimited deterministic signals and the  literature  on estimating continuous-time random processes from their samples  (known in various communities under different names, such as Wiener-Kolmogorov filtering, Gaussian process regression, and kriging).  This leads to the realization that Shannon interpolation can be interpreted as implicitly expecting that the unknown signal has uniform spectral density within its bandwidth.  However, in many practical applications, we often expect more energy at some frequencies than at others.   This leads us to propose novel Shannon-like interpolators that are optimal with respect to  appropriately-constructed weighted Hilbert spaces, where weighting enables us to accommodate prior information about nonuniform spectral density. Although our new interpolants are derived for data obtained with any sampling rate, we observe that they are particularly useful for interpolating sub-Nyquist data.  In addition to theory, we also discuss aspects of practical implementation and show illustrative examples to demonstrate the potential benefits of the approach.
\end{abstract}

\section{Introduction}

The celebrated Sampling Theorem\footnote{This theorem is frequently associated with the names Shannon and Nyquist in the signal processing literature, although many others (including Whittaker, Kotel'nikov, etc.) played important roles  in the development -- readers interested in the historical details are referred to Refs.~\cite{shannon1949,jerri1977,   unser2000, marks1991, meijering2002, butzer2011}.}  states that a bandlimited continuous-time signal $x(t)$ can be perfectly recovered from its uniform samples $x[n]= x(nT)$ for $n\in\mathbb{Z}$, as long as the sampling interval $T$ (with $T>0$) is sufficiently small. This is made explicit by the Shannon interpolation formula (also called the cardinal series expansion, among other names):
\begin{equation}
	x(t)  = \sum_{n=-\infty}^\infty x[n] \mathrm{sinc}\left(t/T - n\right)\label{eq:interp}
\end{equation}
with $\mathrm{sinc}(t) \triangleq \sin(\pi t)/(\pi t)$, 
which reconstructs $x(t)$ perfectly whenever $T$  satisfies the Nyquist criterion $T< \frac{1}{2B}$. Here, $B$ is an upper bound on the maximum frequency component of the signal so that $x(t)$ can be represented via inverse Fourier transform (restricted to the known frequency support) as
\begin{equation}
	x(t) = \frac{1}{2\pi} \int_{-2\pi B}^{2\pi B} X(\Omega) e^{j \Omega t} d\Omega.
\end{equation}

The interpolation formula from Eq.~\eqref{eq:interp} is one of the cornerstones of modern signal processing, communications, and information theory \cite{shannon1949, jerri1977, unser2000}, and  has had a profound impact on our modern world.  This is, in part, because most continuous-time signals of interest (e.g., speech signals, brain signals, communications signals, images, videos, etc.) are roughly bandlimited or can be rendered as such with appropriate use of antialiasing filters, and because the Sampling Theorem allows us to easily manipulate continuous-time signals via sampled discrete-time representations (as needed by digital systems).  

In real applications, it is not generally  practical to measure the infinite number of samples required by Eq.~\eqref{eq:interp}, and it is instead common to employ a truncated summation that only uses a finite number  of consecutive samples:
\begin{equation}
	\hat{x}(t)  = \sum_{n=-N }^{N} x[n] \mathrm{sinc}\left(t/T - n\right),\label{eq:interpt}
\end{equation}
where we have assumed without loss of generality that a total of $2N+1$ samples were acquired centered about $n=0$.  

In this work, we observe that interpolation using Eq.~\eqref{eq:interpt} has an intimate connection to the Lebesgue $\mathcal{L}_2$-norm over a fixed frequency bandwidth (determined by $T$). The use of the Lebesgue measure (which weights every frequency equally) can be viewed as an implicit expectation that a signal's energy could be  distributed uniformly over its bandwidth. 

However, this stands in stark contrast to the signals we encounter in the real world, which often have  more energy at some frequencies than at others.  For instance, many natural signals  have less energy at higher frequencies, e.g., the well-known $1/f$-distribution \cite{keshner1982}. Signals from the same application domain are also often observed to possess fairly consistent spectral density characteristics -- this is true, e.g., for speech signals \cite{dunn1940}, natural images \cite{burton1987}, electrophysiological recordings \cite{freeman2003}, etc.     Finally, even if no spectral density information happens to be available a priori, it is often possible to estimate such information in a signal-dependent way  \cite{robinson1982,marple1987,stoica2005}.

Notably, we also observe that there exist other interpolators  for a different setting (i.e., estimating continuous-time random processes from their samples\footnote{The literature on estimating random processes  is extensive, and goes by different names in different research communities (including Wiener-Kolmogorov filtering, kriging, and Gaussian process regression)  \cite{wiener1949, kolmogorov1941,  cressie1993, rasmussen2006}, with intimate connections to kernel interpolation methods \cite{wahba1990,wendland2010,  scheuerer2011, fasshauer2011,kanagawa2018}.}) that are more capable of incorporating prior information.  This insight leads us to propose a new deterministic alternative to the truncated Shannon interpolation formula Eq.~\eqref{eq:interpt} (applicable even if the Nyquist criterion is violated) that fuses concepts and features of both deterministic and stochastic interpolation.  Our approach is specifically based on replacing the fixed-bandwidth $\mathcal{L}_2$-norm  with an arbitrary-bandwidth weighted Hilbert space norm, where the use of weights can be used to improve accuracy in the presence of prior spectral density information.  Our approach can be viewed as a version of reproducing kernel Hilbert space (RKHS) interpolation \cite{wahba1990, wendland2010, scheuerer2011, fasshauer2011, kanagawa2018}, although the Shannon-like setting means that our kernel construction is distinct  (and potentially more intuitive and easily tailored to specific problems) compared to typical RKHS interpolators.

This paper is organized as follows.  In Sec.~\ref{sec:optim} we discuss some of the well-known and potentially not-so-well known properties of truncated Shannon interpolation, including both deterministic and stochastic perspectives.  The discussion of the deterministic case makes connections to a fixed-bandwidth $\mathcal{L}_2$-norm, while the discussion of the stochastic case makes clear that an alternative interpolation formula might be preferred over truncated Shannon interpolation if a bandlimited signal is known to have nonuniform spectral density.  Motivated by these observations, Sec.~\ref{sec:weight} derives and characterizes new interpolators that possess optimality properties with respect to weighted Hilbert spaces. We then discuss practical implementation considerations and show illustrative examples that demonstrate the potential advantages of the new interpolators in Sec.~\ref{sec:examp}. These examples suggest that these interpolators may be especially advantageous when the Nyquist criterion is violated.  Discussion and conclusions are given in Sec.~\ref{sec:disc}.

As we were finalizing the draft of this paper, we came across previous work \cite{byrne1984,diethorn1991} on different-but-related problems (spectrum estimation/bandlimited signal extrapolation with prior information) that use some of the same weighting concepts that we do.  Of course, there are also many differences because the settings are distinct and the developments were independent and separated by decades.  Nonetheless, readers who enjoy this paper may also appreciate  this earlier work.

\section{Optimality of Truncated Shannon Interpolation (Eq.~\eqref{eq:interpt})}\label{sec:optim}
\subsection{Optimality under a fixed-bandwidth $\mathcal{L}_2$-norm}\label{sec:l2}
The resemblance between the truncated formula Eq.~\eqref{eq:interpt} and the perfect reconstruction formula Eq.~\eqref{eq:interp} should be obvious even  to a casual observer, and this is often used as heuristic justification for the practical utilization of Eq.~\eqref{eq:interpt}.  However, it is perhaps not as widely-known that the signal $\hat{x}(t)$ obtained from Eq.~\eqref{eq:interpt} can also be justified more rigorously as an optimal interpolation of $x[n]$ under multiple different optimality criteria involving the $\mathcal{L}_2$-norm.  For example:
\begin{itemize}
	\item Consider the set $\Gamma_B$ of all continuous-time signals $y(t)$ that perfectly interpolate the measured samples (i.e., $y(nT) = x[n]$ for $-N\leq n \leq N$) and are also $B$-bandlimited, meaning that there exists $Y(\Omega)$ supported on $\Omega \in [-2\pi B,2\pi B]$ such that
	\begin{equation}
y(t) = \frac{1}{2\pi} \int_{-2\pi B}^{2\pi B} Y(\Omega) e^{j \Omega t} d\Omega.
	\end{equation}
Further, let $\Gamma_\star$ denote the special case of $\Gamma_B$ where $T$ and $B$ form a Nyquist pair such that $B = \frac{1}{2T}$.  In this scenario, both the true continuous signal $x(t)$ and the truncated interpolant $\hat{x}(t)$ from Eq.~\eqref{eq:interpt} should be elements of $\Gamma_\star$.  Unfortunately, the set  $\Gamma_\star$ is  uncountably infinite, meaning that there are infinitely many bandlimited signals besides $\hat{x}(t)$ that also perfectly match the measured data samples.  This represents a fundamental ambiguity that cannot be overcome without additional prior information.  Despite this ambiguity, the signal $\hat{x}(t)$ from Eq.~\eqref{eq:interpt} can be shown to possess a certain form of optimality that might make it preferable to other candidates from $\Gamma_\star$.  Specifically, $\hat{x}(t)$ from Eq.~\eqref{eq:interpt} is the unique element of $\Gamma_\star$ with smallest energy, i.e.,
\begin{equation}
	\hat{x}(t) = \arg \min_{y(t) \in \Gamma_\star} \int_{-\infty}^\infty |y(t)|^2 dt.\label{eq:mn}
\end{equation}
This result can be derived using either classical signal processing arguments with minimal assumptions (e.g., \cite[Sec.~6.2.2.2]{liang2000}), or by formulating a minimum-norm  problem in an appropriately-constructed Hilbert space \cite{yao1967,bertero1985,bresler2008a, vandewalle2000}. In particular, Parseval's theorem means that Eq.~\eqref{eq:mn}  can be viewed as a minimum-norm problem in the Hilbert space of bandlimited finite-energy signals  with the following frequency-domain fixed-bandwidth $\mathcal{L}_2$-norm 
\begin{equation}
\|x(t)\|_{\mathcal{L}_2} \triangleq \sqrt{\frac{1}{2\pi}\int_{-\frac{\pi}{T}}^{\frac{\pi}{T}} | X(\Omega)|^2 d\Omega}\label{eq:l2}
\end{equation}
and corresponding fixed-bandwidth $\mathcal{L}_2$-inner product
\begin{equation}
\left<x(t), y(t) \right>_{\mathcal{L}_2} \triangleq \frac{1}{2\pi} \int_{-\frac{\pi}{T}}^\frac{\pi}{T} X(\Omega) Y^*(\Omega) d\Omega,\label{eq:l2ip}
\end{equation}
where $^*$ denotes complex conjugation.

The form of optimality in Eq.~\eqref{eq:mn} suggests that $\hat{x}(t)$ from Eq.~\eqref{eq:interpt} could be viewed as the simplest of all possible bandlimited interpolations of the data (if one were to use the $\mathcal{L}_2$-norm as a measure of complexity), which might therefore be considered as the best bandlimited interpolant candidate if one subscribes to  Occam's Razor.  
\item In addition to knowing that the true signal $x(t)$ should belong to  $\Gamma_\star$ as defined above, we might also possess information that the true $x(t)$ should obey an explicit energy constraint of the form
\begin{equation}
	\int_{-\infty}^\infty |x(t)|^2 dt \leq E^2
\end{equation}
for some known value of $E \in \mathbb{R}$. Note that this can also be expressed in terms of the fixed-bandwidth $\mathcal{L}_2$-norm as $\|x(t) \|_{\mathcal{L}_2} \leq E $.  Let $\Xi$ denote the subset of signals that satisfy this energy constraint.  It can be shown (based on the geometry of $\Gamma_\star$ and the corresponding orthogonality properties of minimum-norm  solutions in Hilbert spaces \cite{luenberger1969,bresler2008a,moon2000} -- see \cite{bresler2008a, barton2007} for detailed arguments) that the signal $\hat{x}(t)$ from Eq.~\eqref{eq:interpt} is also a minimax solution that minimizes the worst-case mean-squared ($\mathcal{L}_2$) interpolation error among all candidate solutions in the intersection of $\Gamma_\star$ with $\Xi$.  Mathematically, we have:
\begin{equation}
\begin{split}
\hat{x}(t) &= \mathop{\mathrm{arg\,min}}_{y(t) \in \Gamma_\star \cap \Xi } \max_{z(t) \in \Gamma_\star \cap \Xi } \int_{-\infty}^\infty |y(t) -z(t) |^2 dt\\
&= \mathop{\mathrm{arg\,min}}_{y(t) \in \Gamma_\star \cap \Xi } \max_{z(t) \in \Gamma_\star \cap \Xi }  \| y(t) - z(t) \|_{\mathcal{L}_2}.
\end{split}\label{eq:minimax}
\end{equation}

\item If we know that $x(t) \in \Gamma_\star \cap \Xi$ as above, then RKHS concepts \cite{yao1967,wendland2010,fasshauer2011,scheuerer2011} (cf. the ``power function" \cite{wendland2010,fasshauer2011,scheuerer2011}) allow us to also precisely characterize the pointwise error associated with the signal $\hat{x}(t)$ from Eq.~\eqref{eq:interpt}, resulting in, e.g., \cite[Eq.~(11.12)]{wendland2010}
\begin{equation}\begin{split}
\!\!\!\left| \hat{ x}(t) - x(t)\right| \leq \frac{C}{\sqrt{T}}\sqrt{1- \sum_{n=-N }^{N} \left| \mathrm{sinc}\left(t/T - n\right)\right|^2},\label{eq:pointwisebound}
\end{split}
\end{equation}
with
\begin{equation}
\begin{split}
\!C \triangleq \sqrt{E^2-  T \sum_{n=-N}^N |x[n]|^2} =  \sqrt{E^2-  \| \hat{x}(t) \|_{\mathcal{L}_2}^2},
\end{split}
\end{equation}
where this result is derived based on the RKHS associated with the fixed-bandwidth $\mathcal{L}_2$-norm.  
It can be further shown \cite{scheuerer2011} (see Appendix~\ref{app:minmaxpointwise} for an intuitive proof that does not require RKHS machinery) that this pointwise error bound is also worst-case optimal, in the sense that for fixed $t$, we have $\hat{x}(t)$ as the solution to the pointwise minimax problem
\begin{equation}
\hat{x}(t) =  \mathop{\mathrm{arg\,min}}_{y(t) \in \Gamma_\star \cap \Xi } \max_{z(t) \in \Gamma_\star \cap \Xi } | y(t) - z(t)|.\label{eq:minmax2}
\end{equation}

\end{itemize}

Together, these deterministic characterizations suggest that $\hat{x}(t)$ is not only the simplest bandlimited interpolation of $x[n]$ (in an appropriate sense associated with the fixed-bandwidth $\mathcal{L}_2$-norm), but it is also the safest (again in appropriate senses associated with the fixed-bandwidth $\mathcal{L}_2$-norm), which may be encouraging and comforting for those choosing to use Eq.~\eqref{eq:interpt}.

Despite these compelling optimality characteristics, the interpolant from Eq.~\eqref{eq:interpt} also has some potentially undesirable features. First, it inflexibly assumes $B=\frac{1}{2T}$ and does not accommodate other bandwidth choices.  Second, the interpolant should be expected to produce substantial aliasing artifacts if the Nyquist condition is violated, and cannot adapt to any additional prior information that might be available about the spectral density of $x(t)$.  Finally, even if the Nyquist condition were satisfied such that $\hat{x}(t)$ is guaranteed to converge pointwise to $x(t)$ in the limit as $N \rightarrow +\infty$,  it is  well-known that this convergence is slow, with potentially major errors between $x(t)$ and $\hat{x}(t)$ when $N$ is not large enough \cite{yao1966, papoulis1966, brown1969, beutler1976,  butzer1982}.

\subsection{Stochastic Optimality}\label{sec:rp}
While deterministic constructions of Shannon interpolation are widely disseminated through introductory signal processing textbooks, there is perhaps not as much awareness that the exact same  formula Eq.~\eqref{eq:interpt} can also be derived under very different assumptions from  a random process perspective.  In particular, Eq.~\eqref{eq:interpt} also emerges as the   formula for  the linear minimum-mean-squared error  (LMMSE) estimator of a wide-sense stationary continuous-time random process from its discrete samples, under specific spectral density assumptions:

\begin{itemize}
\item Let $x(t)$ be a zero-mean wide-sense-stationary random process with ``bandlimited" ($B=\frac{1}{2T}$) power spectral density 
\begin{equation}
	S(\Omega) = \left\{ \begin{array}{ll} \gamma^2, & |\Omega| < \frac{\pi}{T} \\ 0, & \mathrm{else}. \end{array}\right.\label{eq:psd}
\end{equation}
Then $\hat{x}(t)$ from Eq.~\eqref{eq:interpt} is the LMMSE estimator  of $x(t)$, is unbiased (i.e., $E[\hat{x}(t) - x(t)] =0)$, has minimum variance among all unbiased linear estimators, and has  minimum mean-squared error (MMSE) (i.e., minimum $E\left[|\hat{x}(t) - x(t)|^2\right]$) within the class of all linear estimators.  Furthermore, if $x(t)$ is further assumed to be a Gaussian random process, then the LMMSE estimator is also the MMSE estimator with respect to the broader class of potentially-nonlinear estimators. These results are easily obtained as special cases of the general theory of LMMSE estimation for continuous-time random processes with known power spectral densities \cite{wiener1949, kolmogorov1941, cressie1993,rasmussen2006}.
\end{itemize}

Notably, the stochastic optimality of Eq.~\eqref{eq:interpt} only applies when assuming  uniform power spectral density as in Eq.~\eqref{eq:psd} and using Nyquist-rate sampling. Making different spectral density assumptions or different bandwidth assumptions will result in different LMMSE interpolators. Specifically, if we had instead assumed a bandlimited random process model with nonuniform power spectral density and arbitrary bandwidth $B$ that is independent from $T$:
\begin{equation}
	S(\Omega) = \left\{ \begin{array}{ll} |H(\Omega)|^2, & |\Omega| < 2\pi B \\ 0, & \mathrm{else}, \end{array}\right.
\end{equation}
then we would end up with the different LMMSE interpolation formula \cite{cressie1993, rasmussen2006}:
\begin{equation}
	\hat{x}_w(t) = \sum_{n=-N}^N c_n R(t -n T),\label{eq:interp2}
\end{equation}
where $R(\tau)$ is the autocorrelation function (obtained as the inverse Fourier transform of the power spectral density $S(\Omega)$), and the coefficients $c_n$ are selected to ensure that $\hat{x}_w(nT) = x[n]$ for $n=-N,\ldots,N$. 

If a random process is bandlimited and sampled at the Nyquist rate but nothing else is known, then assuming a uniform power spectral density as in Eq.~\eqref{eq:psd} might still be reasonable --  this could be viewed as adopting a noninformative prior that  is noncommittal with regard to missing spectral density information, in alignment with the philosophy of maximum entropy methods \cite{jaynes1957}.  On the other hand, if spectral density information happens to be available and/or if $T$ does not perfectly correspond to the Nyquist rate, then use of the truncated Shannon interpolation formula Eq.~\eqref{eq:interpt} instead of the optimal LMMSE estimator Eq.~\eqref{eq:interp2} should be expected to yield suboptimal performance.

\subsection{Perspectives}
It is important to observe that while Eq.~\eqref{eq:interpt} can emerge as an optimal interpolator under both deterministic and stochastic problem formulations, these scenarios are based on very different assumptions.  The $\mathcal{L}_2$ results from Sec.~\ref{sec:l2} were based on the assumption that $x(t)$ was deterministic with finite energy (i.e., finite $\mathcal{L}_2$-norm) and a well-defined Fourier transform, while the LMMSE results from Sec.~\ref{sec:rp} were based on the assumption that $x(t)$ is a wide-sense stationary random process with a uniform power spectral density over its bandwidth (whose sample functions will generally have infinite energy and Fourier transforms that will not exist).  

Despite these differences, the fact that the optimal LMMSE interpolator yields different interpolation results depending on the sampling rate and/or prior spectral density information in the stochastic setting suggests the possibility of developing new interpolators for the deterministic setting that have similar characteristics.\footnote{See \cite[Sec.~1.4]{wahba1990},\cite{scheuerer2011,kanagawa2018} for further discussion of the parallels between deterministic and stochastic estimation problems. }  This motivates our derivation of new deterministic interpolation formulae in the next section. 

\section{Interpolation in Weighted Hilbert Spaces}\label{sec:weight}
Instead of the fixed-bandwidth $\mathcal{L}_2$-norm and inner product from Eqs.~\eqref{eq:l2} and \eqref{eq:l2ip}, we now consider optimal interpolation in the Hilbert space $\mathcal{W}$ of arbitrary-bandwidth bandlimited continuous-time functions with weighted norm
\begin{equation}
\|x(t)\|_\mathcal{W} \triangleq \sqrt{\frac{1}{2\pi}\int_{-2\pi B}^{2\pi B} | X(\Omega)|^2 W(\Omega) d\Omega}\label{eq:wl2}
\end{equation}
and inner product 
\begin{equation}
\left<x(t), y(t) \right>_{\mathcal{W}} \triangleq \frac{1}{2\pi} \int_{-2\pi B}^{2\pi B} X(\Omega) Y^*(\Omega) W(\Omega) d\Omega.
\end{equation}
For the sake of simplicity, we will assume in this work\footnote{Our restrictions on $W(\Omega)$ are perhaps stronger than they need to be, although weakening them would require a higher degree of mathematical sophistication and make this article harder to understand.  } that $W(\Omega)$ is real-valued, nonnegative, and satisfies $0 < L< W(\Omega) < U < \infty$ for almost every $\Omega \in [-2\pi B, 2\pi B]$ for some lower- and upper-bounds $L, U \in \mathbb{R}$.  Note that this weighted Hilbert space norm can be viewed as a generalization of the fixed-bandwidth $\mathcal{L}_2$-norm from Eq.~\eqref{eq:l2}, and we recover Eq.~\eqref{eq:l2} in the special case where $B= \frac{1}{2T}$ and we apply uniform weighting $W(\Omega)=1$ for $\forall\Omega$ within the bandwidth.

In what follows, we will be interested in deriving a minimum-norm data-consistent interpolant $\hat{x}_w(t)$ with respect to this $\mathcal{W}$-norm, i.e.,
\begin{equation}
\begin{split}
\hat{x}_w(t) = & \mathop{\mathrm{arg\,min}}_{x_w(t) \in\mathcal{W}} \|x_w(t)\|_\mathcal{W} \\ & \text{ s.t. }x_w(nT) = x[n] \text{ for } n=-N,\ldots,N,
\end{split} \label{eq:wmn}
\end{equation}
which can be viewed as a generalization of the minimum $\mathcal{L}_2$-norm optimality criterion from Eq.~\eqref{eq:mn}, where the introduction of $W(\Omega)$ allows us more control over the weight given to different frequency components.  Notably, unlike Sec.~\ref{sec:l2}, these new results are valid for arbitrary sampling rates and  do not assume the Nyquist criterion is satisfied. In the following subsections, we first derive a closed-form expression for $\hat{x}_w(t)$ and  discuss some of its optimality characteristics, before discussing the practical selection of $W(\Omega)$. 

\subsection{Derivation of $\hat{x}_w(t)$}\label{sec:xw}
Minimum-norm solutions in Hilbert spaces are well-understood, and our approach to deriving the minimum $\mathcal{W}$-norm solution follows classical arguments \cite{luenberger1969,moon2000,bresler2008a}. It is well known that minimum-norm solutions in Hilbert spaces, when they exist, can be expressed in terms of adjoint operators \cite{luenberger1969,moon2000,bresler2008a}.  As such,  this section begins by defining point evaluation functionals, which are then used to construct the sampling operator that produces $x[n]$ from $x(t)$, before deriving the adjoint of this sampling operator.

The point evaluation functional for an arbitrary signal $x(t) \in \mathcal{W}$ is linear, continuous, and bounded, and can be represented (per the Riesz representation theorem) via the $\mathcal{W}$ inner product  as
\begin{equation}
\begin{split}
x(t_0)&= \frac{1}{2\pi} \int_{-2\pi B}^{2\pi B} X(\Omega) e^{j \Omega t_0} d\Omega\\
&= \frac{1}{2\pi} \int_{-2\pi B}^{2\pi B} X(\Omega) \left( \frac{e^{-j \Omega t_0}}{W(\Omega)} \right)^* W(\Omega) d\Omega\\
&= \left<x(t), \phi_{t_0}(t) \right>_{\mathcal{W}},
\end{split}
\end{equation}
where
\begin{equation}
\phi_{t_0}(t) \triangleq  \frac{1}{2\pi} \int_{-2\pi B}^{2\pi B} \frac{e^{-j \Omega (t_0-t)}}{W(\Omega)}  d\Omega .
\end{equation}
Note that if we define $\psi(t)$ as the inverse Fourier transform of $1/W(\Omega)$:
\begin{equation}
\psi(t) = \frac{1}{2\pi} \int_{-2\pi B}^{2\pi B} \frac{e^{j \Omega t}}{W(\Omega)}  d\Omega ,\label{eq:psi}
\end{equation}
then we have the simple relationship  $\phi_{t_0}(t) = \psi(t-t_0)$.

Given this representation of point evaluation, we can define the sampling operator $\mathcal{S}: \mathcal{W} \rightarrow \mathbb{C}^{2N+1}$ that generates $x[n]$ from $x(t)$ as
\begin{equation}
[\mathcal{S} \left\{ x(t) \right\}]_n \triangleq \left<x(t), \psi(t-nT) \right>_{\mathcal{W}} 
\end{equation}
for $n=-N,\ldots, N$.

Deriving the adjoint of $\mathcal{S}$ requires us to imbue $\mathbb{C}^{2N+1}$ with Hilbert space structure, and in this work, we use the standard $\ell_2$-norm for this space (corresponding to the use of the  dot product as the inner product).  With this choice, the adjoint $\mathcal{S}^\dagger: \mathbb{C}^{2N+1} \rightarrow \mathcal{W}$ of the sampling operator $\mathcal{S}$ can be derived via the relationship \cite{luenberger1969,moon2000,bresler2008a}
\begin{equation}
\left< \mathcal{S}\{x(t) \}, \mathbf{z} \right>_{\ell_2} = \left< x(t), \{ \mathcal{S}^\dagger \mathbf{z}\}(t) \right>_\mathcal{W}
\end{equation}
for $\forall x(t) \in \mathcal{W}$ and $\forall \mathbf{z} \in \mathbb{C}^{2N+1}$.  It is straightforward to show that this is uniquely satisfied by
\begin{equation}
\{\mathcal{S}^\dagger \mathbf{z}\}(t)  = \sum_{n=-N}^N z_n \psi(t- nT).
\end{equation}

We can now proceed to find the solution to Eq.~\eqref{eq:wmn}.  Classical Hilbert space results \cite{luenberger1969,moon2000,bresler2008a} require that a minimum-norm solution $\hat{x}_w(t)$ (when it exists) must belong to the range of $\mathcal{S}^\dagger$, meaning that $\hat{x}_w(t)$ should take the form
\begin{equation}
\hat{x}_w(t) = \sum_{n=-N}^N c_n \psi(t- nT),\label{eq:xw}
\end{equation}
where the coefficients $c_n$ should be chosen such that $\hat{x}_w(nT) = x[n]$ for $n=-N,\ldots,N$.  These coefficients can be practically obtained by first constructing a matrix $\mathbf{R} \in \mathbb{C}^{(2N+1)\times(2N+1)}$ with entries
\begin{equation}
[\mathbf{R}]_{m n} = \psi\left(mT - nT\right)\label{eq:R1}
\end{equation}
for $m,n = -N,\ldots, N$.\footnote{Note that we have adopted an atypical matrix indexing scheme to substantially simplify notation, where $m$ and $n$ range from $-N,\ldots,N$. We trust that this will not cause confusion. } Then, assuming $\mathbf{R}$ is invertible,\footnote{Under our assumptions on $W(\Omega)$, the positive definiteness (and therefore invertibility) of $\mathbf{R}$ can be  guaranteed by Bochner's theorem  \cite[Ch.~6]{wendland2010}.} we obtain the optimal $c_n$ via
\begin{equation}
\mathbf{c} = \mathbf{R}^{-1} \mathbf{x},
\end{equation}
where $\mathbf{c} \in \mathbb{C}^{2N+1}$ and $\mathbf{x} \in \mathbb{C}^{2N+1}$ are respectively the vectors of the $c_n$ and $x[n]$ values for $n = -N,\ldots, N$.  

Letting $\mathbf{P} = \mathbf{R}^{-1}$  and using $p_{nm}$ to denote the entries of $\mathbf{P}$ for $n,m = -N,\ldots, N$, we can also rewrite the interpolation formula Eq.~\eqref{eq:xw} more explicitly in terms of $x[n]$ as
\begin{equation}
\hat{x}_w(t) = \sum_{n=-N}^N x[n] u_n(t),\label{eq:xwc}
\end{equation}
where $u_n(t)$ are ``cardinal" functions  (see \cite[Sec.~3.1]{wendland2010}) with
\begin{equation}
	u_n(t) = \sum_{m=-N}^N p_{nm} \psi(t -m T),\label{eq:cardinal}
\end{equation}
which satisfy $u_n(mT) = \delta[n-m]$ for $n,m=-N,\ldots,N$.  Here, $\delta[n]$ denotes the Kronecker delta with  $\delta[n] = 0$ for all $n\in\mathbb{Z}\setminus\{0\}$ and $\delta[0]=1$.

Note that, as should be expected, if $B=\frac{1}{2T}$ and we use uniform frequency weights $W(\Omega)=1$, then $\psi(t) =T^{-1} \mathrm{sinc}(t/T)$ and the weighted Hilbert space interpolator $\hat{x}_w(t)$ from Eq.~\eqref{eq:xw} simplifies to the standard truncated Shannon interpolator $\hat{x}(t)$ from Eq.~\eqref{eq:interpt}.

\subsection{Optimality of $\hat{x}_w(t)$}
Because $\hat{x}_w(t)$ from Eq.~\eqref{eq:xw} is a minimum-norm interpolator, it possesses similar optimality characteristics to those described for Eq.~\eqref{eq:interpt} in Sec.~\ref{sec:l2}, but with respect to the $\mathcal{W}$-norm instead of the $\mathcal{L}_2$-norm, and without requiring that $B =\frac{1}{2T}$ or that the Nyquist criterion is satisfied.  In particular:
\begin{itemize}
\item  $\hat{x}_w(t)$ has the minimum $\mathcal{W}$-norm within the set $\Gamma_B$, and can thus be viewed as the simplest of all possible $B$-bandlimited interpolations of the data if one were to use the $\mathcal{W}$-norm as a measure of complexity (cf. Eq.~\eqref{eq:mn}).
\item If we know that the true signal $x(t)$ obeys a $\mathcal{W}$-norm constraint of the form
\begin{equation}
\|x(t)\|_\mathcal{W}  \leq D,
\end{equation}
and let $\Lambda$ denote the set of all signals that satisfy this constraint, then $\hat{x}_w(t)$ is a minimax solution that minimizes the worst-case $\mathcal{W}$-norm error \cite{bresler2008a, barton2007}:
\begin{equation}
\begin{split}
\hat{x}_w(t) &= \mathop{\mathrm{arg\,min}}_{y(t) \in \Gamma_B \cap \Lambda } \max_{z(t) \in \Gamma_B \cap \Lambda }  \| y(t) - z(t) \|_{\mathcal{W}}
\end{split}
\end{equation}
 (cf. Eq.~\eqref{eq:minimax}).
\item If we know that $x(t) \in \Gamma_B \cap \Lambda$, as above, then RKHS concepts also enable a pointwise error bound  similar to Eq.~\eqref{eq:pointwisebound} but relying on the $\mathcal{W}$-norm instead of the fixed-bandwidth $\mathcal{L}_2$-norm.  Specifically, applying \cite[Eq.~(11.12)]{wendland2010} produces
\begin{equation}\begin{split}
\left| \hat{ x}_w(t) - x(t)\right| \leq  \sqrt{D^2-  \| \hat{x}_w(t) \|_{\mathcal{W}}^2}  P(t),
\end{split}
\end{equation}
with
\begin{equation}
\begin{split}
P^2(t) \triangleq & \psi(0) -2 \sum_{n=-N}^N  \mathrm{Real}\left(u^*_n(t) \psi(nT-t)\right)  \\
&\!\!+ \sum_{n=-N}^N \sum_{m=-N}^N u_m(t) \psi(nT-mT) u_n^*(t).
\end{split}
\end{equation}
\item Pointwise minimax optimality of $\hat{x}_w(t)$  can also be shown \cite{scheuerer2011} (cf. Eq.~\eqref{eq:minmax2}). 
\end{itemize}

These optimality results suggest that $\hat{x}_w(t)$ from Eq.~\eqref{eq:xw} has attractive properties if one agrees with use of the $\mathcal{W}$-norm.  We address the rationale for and selection of appropriate weighting functions in the next subsection.

\subsection{Selection  of $W(\Omega)$}

The results of the previous subsections give formulae to calculate $\hat{x}_w(t)$ and characterize some of its properties.  However, it remains to describe reasonable strategies to select a weight function $W(\Omega)$ that will produce desirable results.  Because there can be some degree of subjectivity involved in this choice depending on the type of prior information that is available, we will describe this from different perspectives, i.e., regularization, filtering, and stochastic perspectives. 

\subsubsection{Regularization Perspectives} It is common in the literature on inverse problems to encounter scenarios where there are infinitely many viable solutions to a given inverse problem, and it is necessary to rely on additional (often subjective) criteria to choose a unique solution from the set of candidates.   This is often done by choosing a solution that minimizes a subjectively-chosen regularization penalty, which is designed to favor some candidate solutions over others based on prior knowledge about the properties of the true signal.  For example, regularization penalties are commonly designed to favor smooth solutions, sparse solutions, nonnegative solutions, or solutions with minimum energy \cite{vogel2002,bertero1998,hansen2010, fessler2010}.  As already described, the use of the truncated Shannon interpolation formula $\hat{x}(t)$ from Eq.~\eqref{eq:interpt} is associated with a fixed-bandwidth $\mathcal{L}_2$-norm regularization penalty on the total signal energy, which may be reasonable in some circumstances.  However, if $x(t)$ is known to have nonuniform spectral energy or if $B \neq \frac{1}{2T}$, a fixed-bandwidth $\mathcal{L}_2$-norm penalty would not prioritize solution candidates that are consistent with that prior information.

If prior information is known about a signal's expected energy distribution, it is possible to design regularization penalties that are consistent with that prior.  For example, this can be achieved by using a weighted $\mathcal{L}_2$-norm, where the weights are designed to strongly penalize energy in unexpected places, while giving no or negligible penalties to energy that appears in expected places.  Let's assume that we are given a real-valued positive function  $Z(\Omega)$ that represents our knowledge about the expected spectral density characteristics of $x(t)$, with large values of $Z(\Omega)$ at frequencies where $|X(\Omega)|^2$ is likely to be large and small values where $|X(\Omega)|^2$ is likely to be small.  In this case, the weighting principles above suggest that a reasonable choice of $W(\Omega)$ might be
\begin{equation}
W(\Omega) = \frac{1}{\theta(Z(\Omega) )} \label{eq:reg}
\end{equation}
for a strictly-increasing function $\theta(\tau)$.  This choice achieves larger values of $W(\Omega)$ when $Z(\Omega)$ is small, and vice versa.  A typical choice of $\theta(\cdot)$ might be $\theta(\tau) = (\tau + \varepsilon)^{p/2-1}$ for some $p>0$ and $\varepsilon>0$, which can, e.g., make the $\mathcal{W}$-norm somewhat resemble the $\mathcal{L}_p$-norm when $Z(\Omega) \approx |X(\Omega)|^2$ and $\varepsilon$ is small \cite{vogel2002, chartrand2008a, nikolova2005, black1996, burrus1994, karlovitz1970}.  While this approach to choosing $W(\Omega)$ is quite general and flexible, the selection of $Z(\Omega)$ and $\theta(\cdot)$ are both still subjective.  The alternative perspectives discussed below  are less general but potentially more specific. 

Before moving on, it is worth mentioning that using weights to reflect prior information or desired behavior is not a new idea.  Previous examples of this have appeared in imaging scenarios where prior knowledge is available about the expected locations of image edges, and weights are used to encourage or discourage edge formation in different spatial regions  \cite{leahy1991,fessler1992,hero1999,haldar2008,haldar2012}.   Similar principles are also frequently employed in the design of digital filters \cite{parks1987}, where larger error weights are used in spectral regions where it is important for the filter error (ripple) to be especially small.

It may also be worth noting that iteratively-reweighted least squares methods appear frequently in applications like robust statistics, compressed sensing, approximation theory, and nonlinear optimization.  In these approaches, data-dependent and iteratively-updated weighting functions are frequently leveraged to make quadratic norms (i.e., weighted versions of $\ell_2$- or $\mathcal{L}_2$-norms)  behave more like other regularization penalties such as $\ell_p$ or $\mathcal{L}_p$-norms, Huber functions, etc. \cite{black1996,nikolova2005,vogel2002, chartrand2008a, rice1968, karlovitz1970, burrus1994, gorodnitsky1997}. In these cases, it is also frequent that the data-driven weighting procedure utilizes smaller weights in signal regions with larger amounts of estimated signal energy and vice versa.

\subsubsection{Filtering Perspectives}  In some scenarios, we might be able to model the unknown signal $x(t)$  as a filtered version of some original $B$-bandlimited signal $g(t)$ that is expected to have uniform spectral density over its bandwidth. In this case, the filter is solely responsible for imparting nonuniform spectral density to $x(t)$.\footnote{Modeling a signal with nonuniform spectral density as a filtered version of a signal with uniform spectral density is also common  for stationary random processes, i.e., the Wold representation \cite[Ch.~3]{box1970}, \cite[Ch.~10]{scharf1991}.}  

Let $H(\Omega)$ denote the frequency response of this linear time-invariant filter such that $X(\Omega)= G(\Omega) H(\Omega)$.  We will use the operator $\mathcal{H}$ to denote this filtering operation such that $x(t) = \mathcal{H} g(t)$.   If we further assume that the values of $|H(\Omega)|$ are  bounded away from 0 and $\infty$ almost everywhere within the bandwidth, then the filter is invertible and we can write $G(\Omega) = X(\Omega)/H(\Omega)$ or $g(t) = \mathcal{H}^{-1} x(t)$.  Notably, when $B=\frac{1}{2T}$, $\| g(t) \|_{\mathcal{L}_2}$ can be equivalently written as $\| g(t) \|_{\mathcal{L}_2}=\| \mathcal{H}^{-1}x(t) \|_{\mathcal{L}_2}$, with $\| \mathcal{H}^{-1}x(t) \|_{\mathcal{L}_2} = \|x(t)\|_{\mathcal{W}}$ if $W(\Omega) = 1/|H(\Omega)|^2$ is used to define the $\mathcal{W}$-norm.  Thus, if one chooses to use the $\mathcal{L}_2$-norm for $g(t)$ (a natural choice given that $g(t)$ has uniform spectral density), then it may also  be natural to employ the $\mathcal{W}$-norm with $W(\Omega) = 1/|H(\Omega)|^2$ for $x(t)$.  Notably, this choice of weights can be viewed as a special case of Eq.~\eqref{eq:reg}, and follows the general  philosophy of applying larger weight values to frequencies  where $|X(\Omega)|$ is expected to be smaller and vice versa.

 Interestingly, this formulation also gives further rationale for the use of $\hat{x}_w(t)$ from Eq.~\eqref{eq:wmn}. Specifically, using change-of-variables principles, Eq.~\eqref{eq:wmn} can be equivalently written as $\hat{x}_w(t) = \mathcal{H} \hat{g}(t)$, where $\hat{g}(t)$ is the minimum $\mathcal{L}_2$-norm solution to the inverse problem that seeks to invert both the filtering and sampling operations:
\begin{equation}
\hat{g}(t) = \arg\min_{g(t)} \|g(t)\|_{\mathcal{L}_2} \text{ s.t. } \mathcal{S}\mathcal{H}g(t) = \mathbf{x}.
\end{equation}

\subsubsection{Stochastic Perspectives}  The stochastic optimality concepts described in Sec.~\ref{sec:rp} can also inspire the selection of an appropriate weight function $W(\Omega)$.  In particular, we can make the deterministic interpolator $\hat{x}_w(t)$ from Eq.~\eqref{eq:xw} identical to the LMMSE interpolation formula for random processes (Eq.~\eqref{eq:interp2})  by choosing weights that are inversely proportional to the power spectral density ($W(\Omega) = 1/S(\Omega)$) for frequencies within the bandwidth.  This also coincides with the basic weighting principles described above, where larger weights are applied in spectral regions where $|X(\Omega)|$ is likely to be smaller and vice versa.  In practice, if direct knowledge of the spectral density is not available, it may be reasonable to heuristically apply standard spectral density estimation methods  \cite{robinson1982,marple1987,stoica2005} to obtain a data-dependent estimate of the spectral distribution, even in scenarios where $x(t)$ is not actually a wide-sense stationary random process.

\section{Practical Implementation Considerations and Illustrative Examples}\label{sec:examp}
\subsection{Implementation Considerations}
\subsubsection{Representation of $W(\Omega)$}
Utilization of the weighted Hilbert space interpolator $\hat{x}_w(t)$ from Eq.~\eqref{eq:xw} requires us to choose $W(\Omega)$, from which we must derive the basis function $\psi(t)$ using Eq.~\eqref{eq:psi} and the matrix $\mathbf{R}$ using Eq.~\eqref{eq:R1}.    In practice, although our interpolation formula is valid for arbitrary weight functions $W(\Omega)$ that comply with the assumptions stated in Sec.~\ref{sec:weight}, it will be convenient if we restrict attention to $W(\Omega)$ that produce simple closed-form expressions for $\psi(t)$ that can be evaluated efficiently.  

While there are different ways to achieve this goal, the implementation we use in  later numerical examples is based on representing $1/W(\Omega)$ as a linear combination of simple elementary functions $K_p(\Omega)$, such that
\begin{equation}
	\frac{1}{W(\Omega)} =  \sum_{p=1}^P d_p K_p(\Omega).
\end{equation}
With this representation, we can easily evaluate $\psi(t)$ as
\begin{equation}
	\psi(t) = \sum_{p=1}^P d_p k_p(t),
\end{equation}
where  $k_p(t)$ are the inverse Fourier transforms of $K_p(\Omega)$.  This expression is especially computationally convenient when $k_p(t)$ is easy to evaluate and when $P$ is  small.

Our later examples use $K_p(\Omega)$ functions that correspond to uniformly-translated B-splines \cite{unser1999}.  This may not always be the best possible choice, but it is nonetheless fairly flexible and convenient because B-splines are compactly-supported (which makes it easy to represent bandlimitedness), can accurately represent arbitrary smooth functions, and have analytic Fourier transforms (useful for producing simple expressions for $\psi(t)$). 

Following Unser \cite{unser1999}, let $\beta^K(\Omega)$ denote the  $K$th-degree centered B-spline, which can be written as the successive convolution of $(K+1)$  zero-degree B-splines $\beta^0(\Omega)$, with
\begin{equation}
	\beta^0(\Omega) = \left\{ \begin{array}{ll} 1, & |\Omega| < \frac{1}{2} \\ 0, & \mathrm{else}.\end{array} \right.
\end{equation}
The zero-degree B-spline is a rectangle function, the first-degree B-spline is a triangle function, and the third-degree B-spline (i.e., the cubic B-spline) is widely used for its smooth interpolation characteristics -- see Ref.~\cite{unser1999} for  details.

Our later examples use the B-spline representation
\begin{equation}
	\frac{1}{W(\Omega)} =  \sum_{m=-M}^M d_m \beta^K\left( \frac{\Omega}{2A} -m\right) + \alpha \beta^0\left(\frac{\Omega}{4\pi B}\right),  \label{eq:Gspline}
	\end{equation}
	where the number of $K$th-degree B-splines $(2M+1)$ is a user-selected parameter, the constant $A$ is chosen based on the support of the B-splines so that they are uniformly spaced and symmetrically cover the whole interval $[-2\pi B, + 2\pi B]$:
	\begin{equation}
		A = \frac{ 2 \pi B}{K+2M+1},
	\end{equation}
the coefficients $d_m$ are chosen to produce the desired weighting function $W(\Omega)$, and the  trailing 0-degree B-spline with coefficient $\alpha$ can be used to ensure that Eq.~\eqref{eq:Gspline} is strictly positive for all $\Omega \in [-2\pi B, +2\pi B]$.
 
With this representation of $W(\Omega)$, standard Fourier transform properties result in the simple expression for $\psi(t)$:
\begin{equation}
	\begin{split}
 	\psi(t) =& \frac{A}{\pi   } \left(\mathrm{sinc}\left(\frac{At}{\pi}\right)\right)^{K+1} \sum_{m=-M}^M d_m e^{j 2A m t} \\
	&+  2 \alpha B \mathrm{sinc}(2Bt).
	\end{split}
\end{equation}

\begin{figure}[t]
\centering
\includegraphics{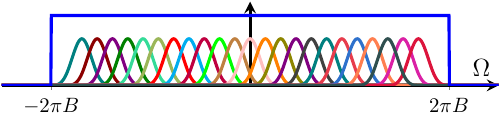}
\caption{The set of B-spline functions used for representing $1/W(\Omega)$ in later numerical examples.}
\label{fig:basis}
\end{figure}

The examples shown later in this paper are all based on the specific choices $K=3$ and $M=11$, with corresponding basis functions shown in Fig.~\ref{fig:basis}.

\subsubsection{Calculation of $\mathbf{R}^{-1}$}
Utilization of the weighted Hilbert space interpolator Eq.~\eqref{eq:xw} requires calculation of $\mathbf{R}^{-1}$.  In the examples shown later in this paper, the number of samples $N$ is small enough that the matrix inversion can be calculated quickly and easily on modern computers.  However, this may no longer be possible if $N$ is very large.  In that case, it may become necessary to rely on iterative methods \cite[Ch.~15]{wendland2010}. As such, it can be useful to observe that $\mathbf{R}$ is both positive definite and Toeplitz, which, e.g., enables the use of fast iterative conjugate gradient solvers  \cite{hestenes1952,chan1996}.  

Another possible computational simplification (which we also do not use in our later examples) is to avoid the full calculation of $\mathbf{R}^{-1}$, and instead focus on the calculation of the cardinal function $u_0(t)$. If one assumes that $u_n(t) \approx u_0(t-nT)$ (which we empirically observe  seems to be reasonably accurate, particularly when $N$ is large and $n$ is near the middle of the sampling interval such that $|n| \ll N$), then a simpler-to-evaluate approximation of Eq.~\eqref{eq:xw} can be obtained with
\begin{equation}
	\hat{x}_w(t)  \approx \sum_{n=-N}^N x[n] u_0(t- nT).
\end{equation}
Note that the coefficients $p_{0m}$ needed to define the cardinal function $u_0(t)$  must uniquely satisfy (see Sec.~\ref{sec:xw})
\begin{equation}
\sum_{m=-N}^N p_{0m} \psi(kT-mT)  = \delta[k] \text{ for } k=-N,\ldots,N.\label{eq:conv}
\end{equation}

Conveniently, this can be viewed  a discrete-time convolution between the unknown coefficients and the samples of $\psi(t)$.  As such, in the limit as $N\rightarrow \infty$, an approximation of the coefficients $p_{0m}$ could be efficiently obtained using Fourier-domain inverse filtering arguments. We leave the details as an exercise -- see \cite[Sec.~II-D]{thevenaz2000} for similar concepts.  

In practice, although $\mathbf{R}$ should be strictly positive definite (and therefore invertible), there are some situations where the matrix is poorly conditioned and the inverse is numerically unstable.  Empirically, we observe this frequently occurs when sampling is faster than the Nyquist rate (i.e., $T < \frac{1}{2B}$), but is less common otherwise.\footnote{Interestingly, this is well-studied for $\psi(t) = \mathrm{sinc}(\alpha t)$. In this case (and consistent with our  empirical observations more generally),  it is known that $\mathbf{R}$  grows increasingly ill-conditioned as $T$ gets smaller  \cite{diethorn1991},\cite[Sec.~10.2.1]{liang2000}.  This behavior is well-known though is perhaps surprising, as one might normally associate denser sampling with an easier interpolation problem. }  While various approaches (e.g., regularization) could be employed to alleviate this issue, such approaches will generally preclude perfect interpolation. Our later examples will focus on cases where $T \geq \frac{1}{2B}$ and matrix inversion is well-behaved.  This does not need to be a limitation in practice, as one could always choose a larger $B$ than necessary (such that $T \geq  \frac{1}{2B}$), while using appropriate weights $W(\Omega)$ to heavily discourage spectral energy in specific frequency ranges.  In addition to bypassing conditioning problems, this approach also imposes bandlimitedness constraints in a softer and gentler way through the combination of $B$ and $W(\Omega)$, instead of imposing strict constraints using $B$ alone.

\subsection{Example I: Signals with Dominant Low Frequencies}
\begin{figure*}
	\begin{center}
		\,\hfill	\subfloat[Weights]{\includegraphics[scale=0.55]{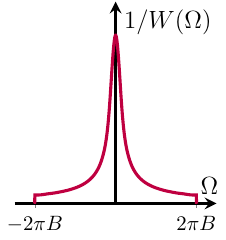}}
		\hfill
		\subfloat[Interpolating function]{\includegraphics[scale=0.55]{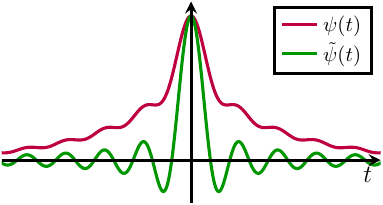}}
		\hfill 
		\subfloat[Nyquist-rate sampling]{\includegraphics[scale=0.55]{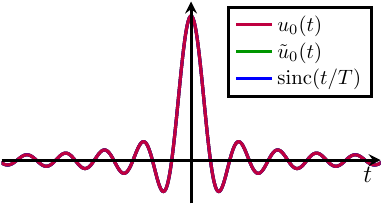}}
		\hfill 
		\subfloat[75\% of Nyquist rate]{\includegraphics[scale=0.55]{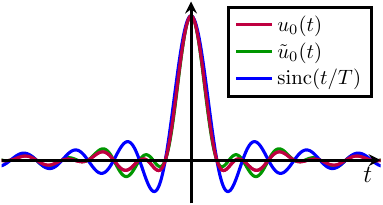}}
		\hfill 
		\subfloat[50\% of Nyquist rate]{\includegraphics[scale=0.55]{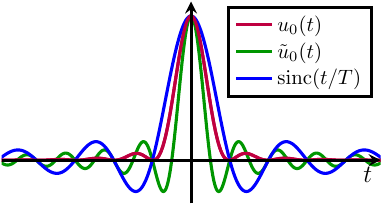}}
		\hfill\,
	\end{center}
	\caption{Example of (a) weights ($1/W(\Omega)$), emphasizing low frequencies, and (b) the  corresponding $\psi(t)$.  For reference, we also show $\tilde{\psi}(t)$, the function  that would have been obtained with uniform weights ($W(\Omega)=1$). In (c)-(d), we show the cardinal functions $u_0(t)$ that are obtained with $N=10$ for different choices of $T$, with comparisons against the cardinal functions $\tilde{u}_0(t)$ that would have been obtained with uniform weights, as well as the $\mathrm{sinc}(t/T)$ functions corresponding to Eq.~\eqref{eq:interpt}.  }\label{fig:lf}
\end{figure*}

Our first illustration considers the weight function shown in Fig.~\ref{fig:lf}(a). These weights might be appropriate for scenarios where spectral energy is expected to be concentrated at low-frequencies, e.g., the $1/f$-distribution.  As can be seen in Fig.~\ref{fig:lf}(b), the interpolating function $\psi(t)$ that will be used to construct $\hat{x}_w(t)$ is substantially smoother and less-oscillatory than the sinc function associated with an unweighted $\mathcal{L}_2$-norm over the same bandwidth (shown as $\tilde{\psi}(t)$).

\begin{figure*}
	\begin{center}
		\,\hfill	
		\subfloat[Spectrum]{\includegraphics[scale=0.51]{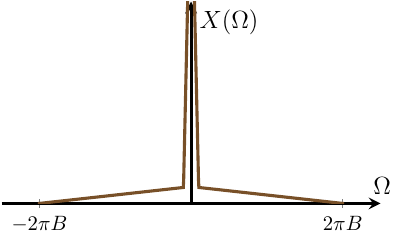}}
		\hfill
		\subfloat[Interpolation with $W(\Omega)$]{\includegraphics[scale=0.51]{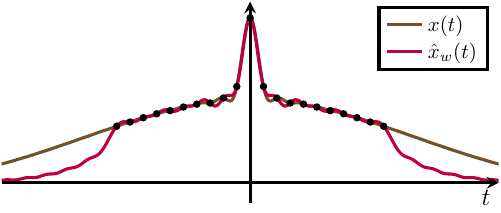}}
		\hfill 
		\subfloat[Interpolation with uniform weights]{\includegraphics[scale=0.51]{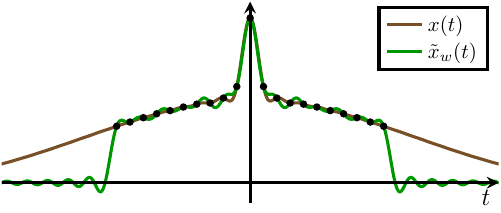}}
		\hfill		\subfloat[Sinc interpolation]{\includegraphics[scale=0.51]{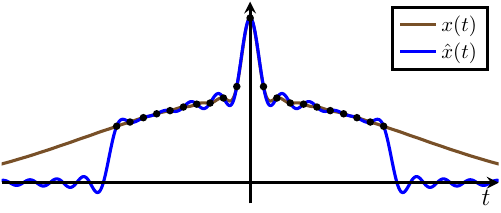}}
		\hfill\,
	\end{center}
\vspace{-2em}
	\begin{center}
		\,\hfill	
		\includegraphics[scale=0.51]{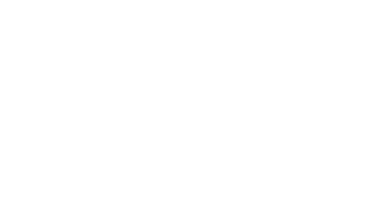}
		\hfill
		\subfloat[Interpolation with $W(\Omega)$]{\includegraphics[scale=0.51]{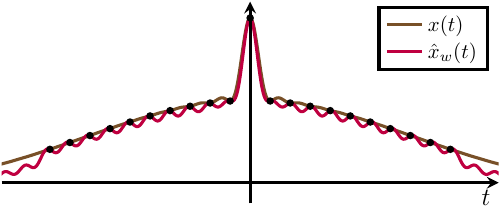}}
		\hfill 
		\subfloat[Interpolation with uniform weights]{\includegraphics[scale=0.51]{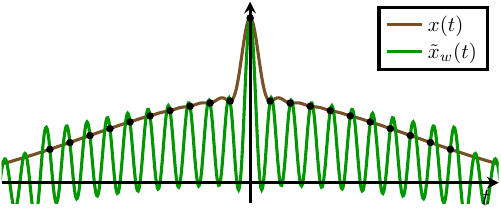}}
		\hfill		\subfloat[Sinc interpolation]{\includegraphics[scale=0.51]{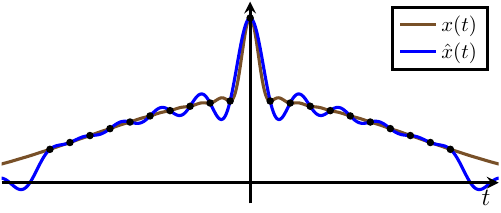}}
		\hfill\,
	\end{center}
	\caption{Interpolating a signal $x(t)$ with the spectrum $X(\Omega)$ as shown in (a), with $N=10$ and the weights from Fig.~\ref{fig:lf}.  We show interpolation results from data acquired at (b-d) 75\% of the Nyquist rate and (e-f) 50\% of the Nyquist rate corresponding to  (b,e) the weighted Hilbert space interpolator $\hat{x}_w(t)$, (c,f) the Hilbert space interpolator $\tilde{x}_w(t)$ corresponding to  uniform weights, and (d,g) the $\mathrm{sinc}(t/T)$ interpolator from Eq.~\eqref{eq:interpt} (ignoring the violation of the Nyquist criterion).}
	\label{fig:lfi}
\end{figure*}

Interestingly, the behavior of the corresponding interpolator depends strongly on the choice of $T$.  As shown in Fig.~\ref{fig:lf}(c), when $T$ is chosen at the Nyquist-rate, the cardinal function $u_0(t)$ of the weighted Hilbert space interpolator is very similar (though still formally distinct) to the interpolation kernel that would be obtained with uniform spectral weights ($W(\Omega)=1$, which results in standard sinc interpolation Eq.~\eqref{eq:interpt}).  This suggests that the use of $\hat{x}_w(t)$ should be expected to be similar standard sinc interpolation when $B=\frac{1}{2T}$.  This is potentially disappointing and perhaps surprising, although becomes less surprising after some additional analysis. Specifically, it is easily proven by taking the discrete-time Fourier transform of the convolution in Eq.~\eqref{eq:conv}  (cf.  \cite{thevenaz2000, unser2000}), that when $B=\frac{1}{2T}$ and for arbitrary $W(\Omega)$ satisfying the conditions of Sec.~\ref{sec:weight}, we must have that $u_0(t) \rightarrow \mathrm{sinc}(t/T)$ in the limit as $N\rightarrow \infty$.  What is perhaps still surprising is that $u_0(t)$ can be very similar to $\psi(t)$  even when $N$ is relatively small.  Although we have observed empirically that $u_n(t)$ starts to diverge more and more from $\mathrm{sinc}(t/T-n)$ as $|n|$ increases, the differences we saw were small, and we would not expect the  new weighted Hilbert space interpolator to have  obvious advantages over Shannon interpolation.  

However, as shown in Fig.~\ref{fig:lf}(d,e), the weighted Hilbert space interpolator and the uniform-weight Hilbert space interpolator have cardinal functions that substantially differ from Shannon interpolation in the sub-Nyquist regime.  In particular, we observe that the cardinal function $u_0(t)$ from the weighted Hilbert space formulation is less oscillatory than either the cardinal function $\tilde{u}_0(t)$ corresponding to the unweighted Hilbert space formulation or the $\mathrm{sinc}(t/T)$ interpolator corresponding to Eq.~\eqref{eq:interpt} (disregarding the fact that the Nyquist criterion has been violated).  As such, we should expect the weighted Hilbert space interpolator to produce smoother results than these alternative interpolants in the sub-Nyquist regime, which is consistent with the prior knowledge that the signal has its energy concentrated at low-frequencies.

\begin{figure*}
	\begin{center}
		\,\hfill	\subfloat[Weights]{\includegraphics[scale=0.55]{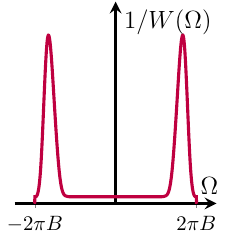}}
		\hfill
		\subfloat[Interpolating function]{\includegraphics[scale=0.55]{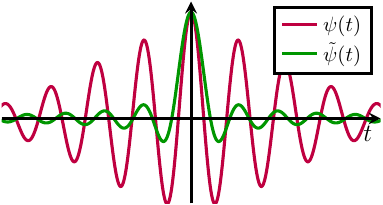}}
		\hfill 
		\subfloat[Nyquist-rate sampling]{\includegraphics[scale=0.55]{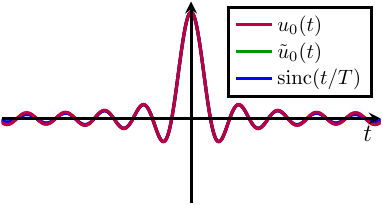}}
		\hfill 
		\subfloat[75\% of Nyquist rate]{\includegraphics[scale=0.55]{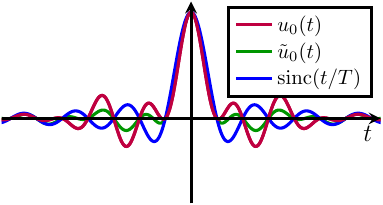}}
		\hfill 
		\subfloat[50\% of Nyquist rate]{\includegraphics[scale=0.55]{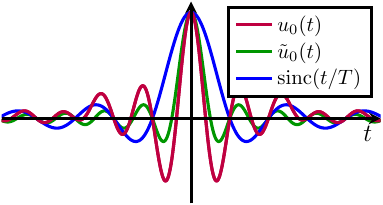}}
		\hfill\,
	\end{center}
	\caption{Example of (a) weights ($1/W(\Omega)$), emphasizing high frequencies, and (b) the  corresponding $\psi(t)$.  For reference, we also show $\tilde{\psi}(t)$, the function  that would have been obtained with uniform weights ($W(\Omega)=1$). In (c)-(d), we show the cardinal functions $u_0(t)$ that are obtained with $N=10$ for different choices of $T$, with comparisons against the cardinal functions $\tilde{u}_0(t)$ that would have been obtained with uniform weights, as well as the $\mathrm{sinc}(t/T)$ functions corresponding to Eq.~\eqref{eq:interpt}.  }\label{fig:hf}
\end{figure*}

\begin{figure*}
	\begin{center}
		\,\hfill	
		\subfloat[Spectrum]{\includegraphics[scale=0.51]{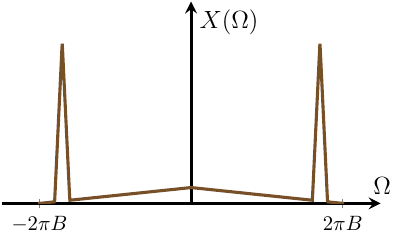}}
		\hfill
		\subfloat[Interpolation with $W(\Omega)$]{\includegraphics[scale=0.51]{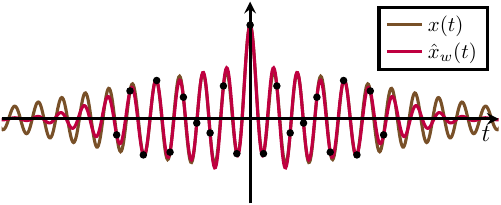}}
		\hfill 
		\subfloat[Interpolation with uniform weights]{\includegraphics[scale=0.51]{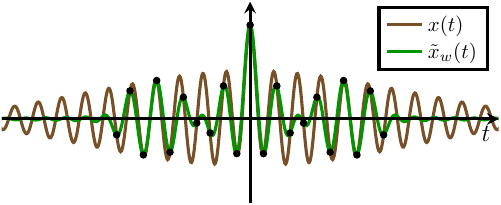}}
		\hfill		\subfloat[Sinc interpolation]{\includegraphics[scale=0.51]{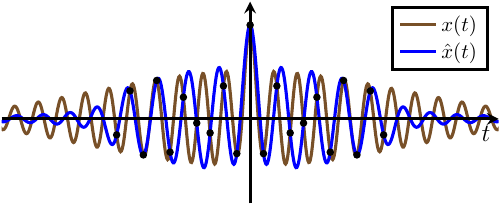}}
		\hfill\,
	\end{center}
	\vspace{-2em}
	\begin{center}
		\,\hfill	
		\includegraphics[scale=0.51]{./empty.pdf}
		\hfill
		\subfloat[Interpolation with $W(\Omega)$]{\includegraphics[scale=0.51]{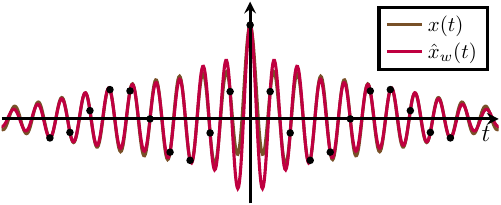}}
		\hfill 
		\subfloat[Interpolation with uniform weights]{\includegraphics[scale=0.51]{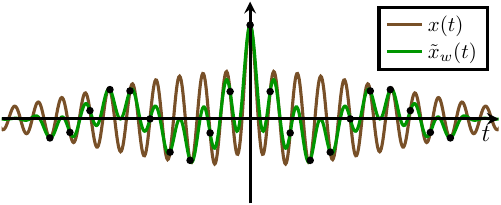}}
		\hfill		\subfloat[Sinc interpolation]{\includegraphics[scale=0.51]{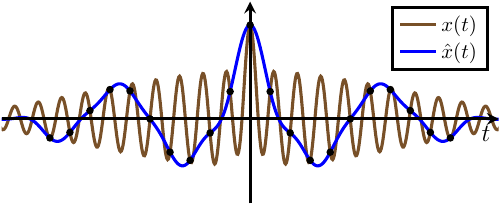}}
		\hfill\,
	\end{center}
	\caption{Interpolating a signal $x(t)$ with the spectrum $X(\Omega)$ as shown in (a), with $N=10$ and the weights from Fig.~\ref{fig:hf}.  We show interpolation results from data acquired at (b-d) 75\% of the Nyquist rate and (e-f) 50\% of the Nyquist rate corresponding to  (b,e) the weighted Hilbert space interpolator $\hat{x}_w(t)$, (c,f) the Hilbert space interpolator $\tilde{x}_w(t)$ corresponding to  uniform weights, and (d,g) the $\mathrm{sinc}(t/T)$ interpolator from Eq.~\eqref{eq:interpt} (ignoring the violation of the Nyquist criterion).}
	\label{fig:hfi}
\end{figure*}

Interestingly, although $u_0(t)$ has infinite support in the time domain, we observe that it can be accurately approximated as having finite-support in the sub-Nyquist regime, particularly when sampling at 50\% of the Nyquist rate $(T=1/B)$.  We do not pursue this here, but such an approximation could be used to substantially simplify computation of the interpolant.

Figure~\ref{fig:lfi} illustrates the behavior when this interpolator is applied to samples of the signal $x(t) = \mathrm{sinc}^2(B t) + \mathrm{sinc}^2(\frac{B}{20} t)$, which is low-frequency dominant.  We omit showing Nyquist-rate interpolation, as all interpolants behave similarly in that case. When sampling at 75\% the Nyquist rate (i.e., $T=\frac{2}{3B}$), we observe that the weighted Hilbert space interpolant is the most accurate while also being the least oscillatory (although in this regime, all three interpolants produce fairly similar results for times $t$ within sampling interval).  However, the differences are much more obvious when sampling at 50\% the Nyquist rate (i.e., $T=\frac{1}{B}$), where the weighted Hilbert space interpolator is fairly accurate, while the Hilbert space interpolator with uniform weights becomes highly oscillatory and very inaccurate and $\mathrm{sinc}(t/T)$ interpolation (ignoring the Nyquist violation) also demonstrates substantial ripple errors.  This case, which is quite difficult because of the Nyquist violation, clearly illustrates the potential value of adopting a weighted Hilbert space approach.

\subsection{Example II: Signals with Dominant High Frequencies}
Our second illustration considers the weight function shown in Fig.~\ref{fig:hf}(a). These weights might be appropriate for scenarios where spectral energy is expected to be concentrated at high-frequencies.  As can be seen in Fig.~\ref{fig:hf}(b), the resulting interpolating function $\psi(t)$ is now substantially more oscillatory  than the sinc function associated with bandwidth $B$.  

Similar to the previous case, the proposed weighted Hilbert space interpolator should be expected to behave similarly to standard Shannon interpolation when data is sampled at the Nyquist rate (as can be seen from the cardinal functions in Fig.~\ref{fig:hf}(c)), although should be expected to demonstrate different behavior in the sub-Nyquist regime (see Fig.~\ref{fig:hf}(d,e)).

Figure~\ref{fig:hfi} illustrates the behavior when this interpolator is applied to samples of the signal $x(t) = \mathrm{sinc}^2(B t) + \mathrm{sinc}^2(\frac{B}{20} t) \cos(1.7 \pi B t)$, which is high-frequency dominant.  As before, we omit showing Nyquist-rate interpolation, as all interpolants behave similarly in that case. As can be seen in Fig.~\ref{fig:hfi} we observe that the weighted Hilbert space interpolator is substantially more accurate than the other interpolators when sampling at both 75\% and 50\% of the Nyquist rate.

\subsection{Example III: Fourier Imaging Data}

\begin{figure}
	\begin{center}
		\includegraphics[width=0.83in]{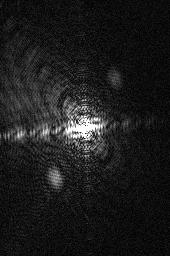}
		\hfill
		\includegraphics[width=0.83in]{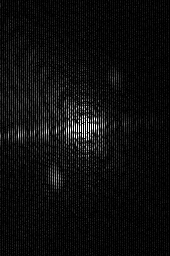}
		\hfill
		\includegraphics[width=0.83in,trim={552px 832px 553px 832px},clip]{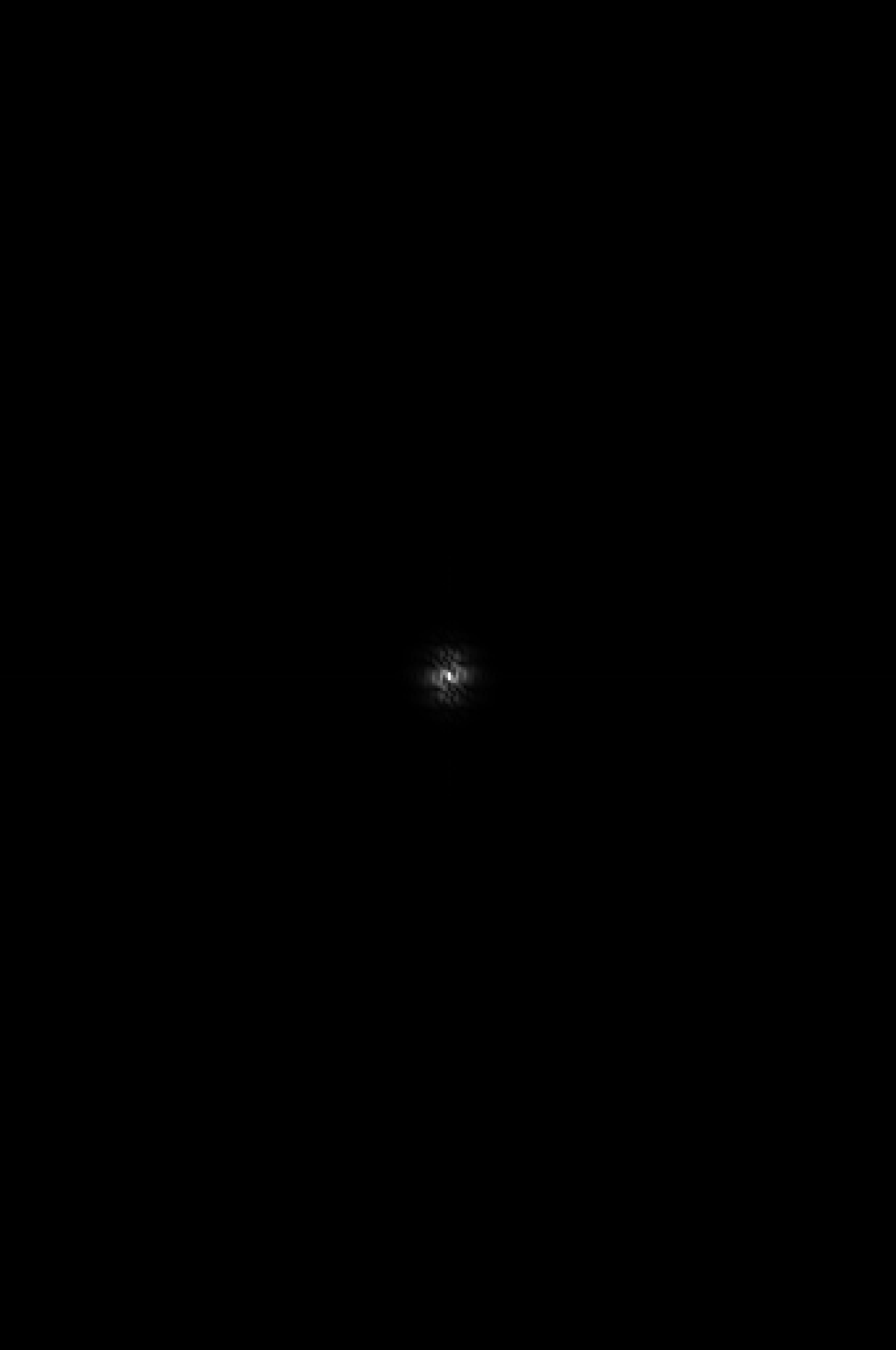}
		\hfill
		\includegraphics[width=0.83in]{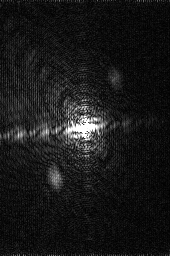}
	\end{center}\vspace{-2em}
	\begin{center}
		\subfloat[Nyquist Data]{\includegraphics[width=0.83in]{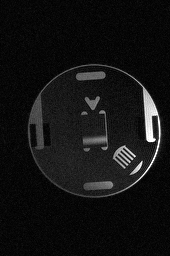}}
		\hfill
		\subfloat[Unweighted]{\includegraphics[width=0.83in]{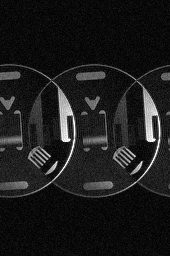}}
		\hfill
		\subfloat[$\psi(\cdot)$ and $\frac{1}{W(\cdot)}$]{\includegraphics[width=0.83in]{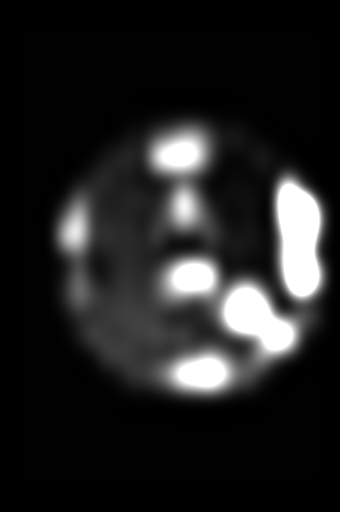}\captionsetup{font=tiny}}
		\hfill
		\subfloat[Weighted]{\includegraphics[width=0.83in]{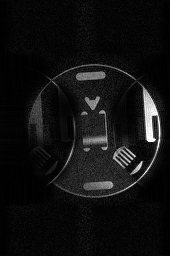}}
	\end{center}
	\caption{Illustration of interpolating MRI data (available from: {\color{blue}https://mr.usc.edu/download/data/}). The top row shows the raw data measurements, while the bottom row shows the corresponding images.  (a) Nyquist-rate data.  (b) Interpolation of  50\%-Nyquist data without weights.  (c) Bottom: weights ($1/W(\cdot)$) and top: $\psi(\cdot)$ (at $\sim5.3\times$ magnification) obtained from prescan data.  (d) Weighted Hilbert space interpolation of 50\%-Nyquist data.}
	\label{fig:mri}
\end{figure}

Our final illustration demonstrates the potential advantages of using  weighted Hilbert space interpolators with real magnetic resonance imaging (MRI) data.  Figure~\ref{fig:mri}(a) shows raw MRI measurements sampled at the Nyquist rate, which correspond to a high-quality image.  If data is sampled at 50\% the Nyquist rate (which offers substantial improvements in data acquisition speed) and interpolated using unweighted Hilbert space interpolation (equivalent to sinc interpolation), we observe a poor interpolation result corresponding to a highly-aliased image, as seen in Fig.~\ref{fig:mri}(b).  However, using a small amount of prescan data (frequently called calibration data in the MRI literature \cite{griswold2002,haldar2020}), it becomes possible to obtain a low-resolution estimate of the image energy distribution (represented by $1/W(\cdot)$), as shown in Fig.~\ref{fig:mri}(c).  The weighted Hilbert space interpolator based on these weights produces the more accurate interpolation results shown in Fig.~\ref{fig:mri}(d).\footnote{Perhaps due to the presence of noise in real data, the interpolation has some unexpectedly high intensities at the boundaries of the sampled data.  These are visible in the interpolated raw data in Fig.~\ref{fig:mri}(d), but have been manually set to zero before computing the corresponding image.}  While the resulting image still contains major errors (as should be expected given the substantial violation of the Nyquist condition), the results in this case are nevertheless much more accurate than those of Fig.~\ref{fig:mri}(b).

\section{Discussion and Conclusion}\label{sec:disc}

In this work, we demonstrated that the use of prior spectral density information can be used to construct novel interpolators that are derived from weighted Hilbert space norms.  These interpolators were demonstrated empirically to enable improved accuracy relative to the conventional truncated Shannon interpolation formula, particularly when sampling below the Nyquist rate. For simplicity, we have presented our results in the context of uniform 1D sampling, although the principles  are general and easily extended to multidimensional \cite[Ch.~6]{marks1991}, \cite{dudgeon1984} and/or nonuniform \cite{marvasti2001} sampling scenarios.

Our interpolators were designed for the noiseless setting, where $x[n]$ is measured exactly and we require that the interpolator $\hat{x}_w(nT)$ must exactly match the samples $x[n]$  for $n=-N,\ldots,N$.  This is easily relaxed to accommodate noisy measurements, and for example, in the presence of i.i.d. zero-mean noise with variance $\sigma^2$, the optimal LMMSE estimator from Eq.~\eqref{eq:interp2} remains optimal as long as $\mathbf{R}^{-1}$ in Eq.\eqref{eq:R1} is replaced with $(\mathbf{R}+ \sigma^2\mathbf{I})^{-1}$ \cite[Sec.~2.2]{rasmussen2006}.  In a deterministic context, this can be viewed as a standard form of (Tikhonov-style) regularization, and could also be a sensible adaptation to use with the weighted Hilbert space interpolator Eq.~\eqref{eq:xw} to avoid ill-conditioning and excessive noise sensitivity.

Although we have framed our contribution broadly for the larger signal processing audience, this work was born out of a desire to address a specific interpolation problem that arises in Fourier imaging contexts like MRI.  Specifically, previous computational imaging work has shown that, under appropriate assumptions, it can be possible to use data-dependent linear prediction/interpolation methods to recover missing Nyquist-rate samples from  sub-Nyquist data \cite{haldar2020, griswold2002, huang2005, lustig2010, shin2014, haldar2013b, haldar2015, ongie2016, jin2015a,lobos2021}, unifying, generalizing, simplifying, and robustifying a range of earlier computational image reconstruction methods \cite{haldar2020,haldar2022}.  While such approaches have proven powerful, they have largely\footnote{See, e.g., Refs.~\cite{seiberlich2007,luo2019} for exceptions.} been formulated for discrete-time\footnote{We use ``time" here loosely -- e.g., in MRI, we are actually interested in interpolating $k$-space, where $k$ corresponds to spatial frequency \cite{liang2000}.} models,  and the genesis of the present effort came from our suspicion that the same concepts could be generalized to enable continuous-time interpolation and provide deeper insights into the underlying structural characteristics of Fourier imaging data.    This led us down some unexpected paths, where we initially delved into the connections between linear predictive modeling and autocorrelation/spectral density modeling \cite[Ch.~3]{box1970},\cite[Ch.~10]{scharf1991}, which then lead us to think about Gaussian process regression and RKHSs, and finally culminated in the present contribution.  We believe that the results reported in this paper represent an important step towards our original goal, although additional developments are still needed  before our continuous-time interpolators possess the same performance levels and suite of capabilities as the existing discrete-time methods for computational imaging applications -- we anticipate that this will be a promising research direction, though defer further explorations to the future.  Members of the MRI image reconstruction community may also be interested to know that, even though the modeling assumptions and methods are quite distinct, there are intriguing commonalities  between our weighted Hilbert space interpolation framework and certain reference-based MRI reconstruction approaches like the generalized series model \cite{liang1994} and k-t BLAST \cite{tsao2003}, suggesting the potential for an even larger-scale unification of diverse existing image reconstruction concepts.

While this work has focused on interpolation in scenarios where we employ fixed weights $W(\Omega)$ based on fixed prior spectral density information, we expect that it may also be fruitful to pursue formulations that perform spectral density estimation and interpolation jointly, potentially using iterative reweighting.  This type of joint approach has proven advantageous in previous discrete-time linear prediction/interpolation problems in imaging where structured low-rank matrix models have been used to adaptively update linear prediction coefficients \cite{haldar2020, shin2014, haldar2013b, haldar2015, ongie2016, jin2015a,lobos2021} (with linear prediction coefficients intimately related to spectral density models \cite[Ch.~3]{box1970},\cite[Ch.~10]{scharf1991}). Similar concepts may be useful in continuous-time.

Our results are based on recognizing parallels between classical deterministic Shannon interpolation and the interpolation of random processes.  In a similar vein, it may be fruitful to view other deterministic interpolation schemes through a stochastic lens.  For example, interpolation with B-splines can be derived deterministically  \cite{unser1999}, but it is perhaps not as well appreciated that there is also a (non-bandlimited!) random process model that will produce  identical interpolants.  This view may be beneficial in designing new interpolation kernels that have desirable practical features (e.g., compact support) while also capturing more of the behavior of  real data.

\appendices

\section{Proof of Pointwise Minimax Optimality (Eq.~\eqref{eq:minmax2})}\label{app:minmaxpointwise}
In this section, we prove Eq.~\eqref{eq:minmax2}, following similar arguments to those that were used in \cite[Sec.~1.8.3]{bresler2008a} to prove  a generalized version of Eq.~\eqref{eq:minimax}.

Note that every finite-energy data-consistent bandlimited interpolant $z(t) \in \Gamma$ can be decomposed as
\begin{equation}
	\begin{split}
	z(t) = & \hat{x}(t)+ \sum_{n\in \Delta  } g[n] \mathrm{sinc}\left(t/T - n\right)
	\end{split}
\end{equation}
for appropriate coefficients $g[n]$ and $\Delta \triangleq \mathbb{Z} \setminus \{-N,\ldots,N\}$.  The further requirement that $z(t) \in \Xi$ implies that
\begin{equation}
\sum_{n\in \Delta } |g[n]|^2 \triangleq \|\mathbf{g}\|^2_{\ell_2} \leq \frac{E^2}{T} - \sum_{n=-N}^N |x[n]|^2 =C^2/T.
\end{equation}
Now, for fixed $t$ and a fixed signal $y(t) \in \Gamma_\star \cap \Xi$, consider the $z(t)$-subproblem from Eq.~\eqref{eq:minmax2}:
\begin{equation}
	\begin{split}
&	\max_{z(t) \in \Gamma_\star \cap \Xi } |y(t) - z(t)|\\
&\!\!\!\!= \max_{\substack{g[n]: \\ \|\mathbf{g}\|^2_{\ell_2} \leq C^2/T }} \left| \sum_{n\in \Delta } g[n] \mathrm{sinc}\left(t/T - n\right) + \hat{x}(t) - y(t)  \right|\\
&\!\!\!\!\leq \max_{\substack{g[n]: \\\|\mathbf{g}\|^2_{\ell_2} \leq C^2/T }} \left| \sum_{n\in \Delta } g[n] \mathrm{sinc}\left(t/T - n\right) \right| + \left| \hat{x}(t)  - y(t)  \right|.
	\end{split}\label{eq:proofminimax}
\end{equation}
Note that the inequality in the last line of Eq.~\eqref{eq:proofminimax}  becomes an equality whenever the phase of $\sum_{n\in \Delta } g[n] \mathrm{sinc}\left(t/T - n\right)$ (which can be selected arbitrarily without changing the value of the upper bound) is chosen to match the phase of $\hat{x}(t)  - y(t)$.  As a result, the upper bound in Eq.~\eqref{eq:proofminimax} is always achievable  and the inequality can be replaced with equality. Thus, returning back to the original problem of Eq.~\eqref{eq:minmax2}, we observe that the minimax optimal $y(t)$ is $y(t) = \hat{x}(t)$.  \qed

Note also that when we make the minimax optimal choice of $y(t) = \hat{x}(t)$, a worst-case $g[n]$ (which depends on the fixed value of $t$) is obtained from matched filtering principles as
\begin{equation}
g[n] = \frac{C}{\sqrt{T}} e^{j\phi} \frac{\mathrm{sinc}\left(t/T - n\right)}{\sqrt{ \sum_{m\in \Delta } |\mathrm{sinc}\left(t/T - m\right)|^2 }} 
\end{equation}
where the phase $\phi$ can be chosen arbitrarily, leading to
\begin{equation}
\begin{split}
\min_{y(t) \in \Gamma \cap \Xi }\max_{z(t) \in \Gamma \cap \Xi } &|y(t) - z(t)|\\
&= \frac{C}{\sqrt{T}} \sqrt{ \sum_{n\in \Delta } |\mathrm{sinc}\left(t/T - n\right)|^2 }.
\end{split}
\end{equation}
This coincides with the upper bound from Eq.~\eqref{eq:pointwisebound} based on the fact that $\sum_{n\in\mathbb{Z}}  |\mathrm{sinc}\left(t/T - n\right)|^2 =1$ for $\forall t$.

\bibliographystyle{IEEEtran}
\bibliography{/home/haldar/spin/002.TeX/000.JabRefBibliography/bibliography}

\end{document}